\documentclass{article}\usepackage{jkas2}
\runningauthor{ENSSLIN}
\runningtitle{Facts and Fiction}
\beginpage{1}
\endpage{5}

\begin{document}

\font\twelvei = cmmi10 scaled\magstep1 
       \font\teni = cmmi10 \font\seveni = cmmi7
\font\mbf = cmmib10 scaled\magstep1
       \font\mbfs = cmmib10 \font\mbfss = cmmib10 scaled 833
\font\msybf = cmbsy10 scaled\magstep1
       \font\msybfs = cmbsy10 \font\msybfss = cmbsy10 scaled 833
\textfont1 = \twelvei
       \scriptfont1 = \twelvei \scriptscriptfont1 = \teni
       \def\mit{\fam1 }
\textfont9 = \mbf
       \scriptfont9 = \mbfs \scriptscriptfont9 = \mbfss
       \def\bmit{\fam9 }
\textfont10 = \msybf
       \scriptfont10 = \msybfs \scriptscriptfont10 = \msybfss
       \def\bmsy{\fam10 }

\def\etal{{\it et al.~}}
\def\eg{{\it e.g.,~}}
\def\ie{{\it i.e.,~}}
\def\lsim{\raise0.3ex\hbox{$<$}\kern-0.75em{\lower0.65ex\hbox{$\sim$}}}
\def\gsim{\raise0.3ex\hbox{$>$}\kern-0.75em{\lower0.65ex\hbox{$\sim$}}}

\def\NoCite#1{}

\title{Extragalactic Cosmic Rays and Magnetic Fields:\\
Facts and Fiction}

\author{Torsten En{\ss}lin}
\address{Max-Planck-Institute for Astrophysics,
Karl-Schwarzschild-Str. 1, 85741 Garching, Germany\\
{\it E-mail: ensslin@mpa-garching.mpg.de}}

%\author{and}
%
%\author{...}
%\address{...}

%\author{...$^{1}$ and ...$^{2}$}
%\address{$^{1}$ ...}
%\address{$^{2}$ ...}

%\author{~}
\address{\normalsize{\it (Received November 15, 2004; Accepted December 1,2004)}}

\abstract{A critical discussion of our knowledge about extragalactic
cosmic rays and magnetic fields is attempted. What do we know for
sure? What are our prejudices? How do we confront our models with the
observations?  How can we assess the uncertainties in our modeling and
in our observations? Unfortunately, perfect answers to these questions
can not be given. Instead, I describe efforts I am involved in to gain
reliable information about relativistic particles and magnetic fields
in extragalactic space.}

\keywords{magnetic fields; cosmic rays; turbulence}

\maketitle

%\tableofcontents

\section {What do we know?}

We know that cosmic rays and magnetic fields exist in clusters of
galaxies for several reasons. First, in many galaxy clusters we
observe the so-called {\it cluster radio halos} with a spatial
distribution which is very similar to that of the intra-cluster gas
observed in X-rays. These radio halos are produced by
radio-synchrotron emitting relativistic electrons (cosmic ray
electrons = CRe) spiraling in magnetic fields. We do not have direct
evidence of cosmic ray protons (CRp) probably due to their much weaker
radiative interactions. However, in our own Galaxy the CRp energy
density outnumbers the CRe energy density by two orders of magnitude,
which makes the assumption of a CRp population in galaxy clusters very
plausible.

Second, the Faraday rotation of linearly polarized radio emission
traversing the intra-cluster medium (ICM) proves independently the
existence of intracluster magnetic fields. It has been debated, if the
magnetic fields seen by the Faraday effect exist on cluster scales in
the ICM, or in a mixing layer around the radio plasma which emits the
polarized emission (Bicknell \etal 1990, Rudnick \& Blundell
1990)\NoCite{1990ApJ...357..373B, 2003ApJ...588..143R}. However, there
is no valid indication of a source local Faraday effect in the
discussed cases (En{\ss}lin \etal 2003)\NoCite{2003ApJ...597..870E},
and the Faraday rotation signal excess of radio sources behind
clusters compared to a field control sample strongly supports the
existence of strong magnetic fields in the wider ICM (Clarke et
al. 2001, Johnston-Hollitt \& Ekers 2004)\NoCite{2001ApJ...547L.111C,
2004astro.ph.11045J}.  The detailed mapping of the Faraday effect of
extended radio sources reveals that the ICM magnetic fields are
turbulent, with power on a variety of scales, and with a
power-spectrum which is Kolmogoroff-type (see Vogt \& En{\ss}lin,
these proceedings). All these Faraday rotation measurements support
magnetic field strengths of the order of several $\mu$G.

\begin {figure*}[t]
%\vskip -0.6cm
\centerline{\epsfysize=11.8cm\epsfbox{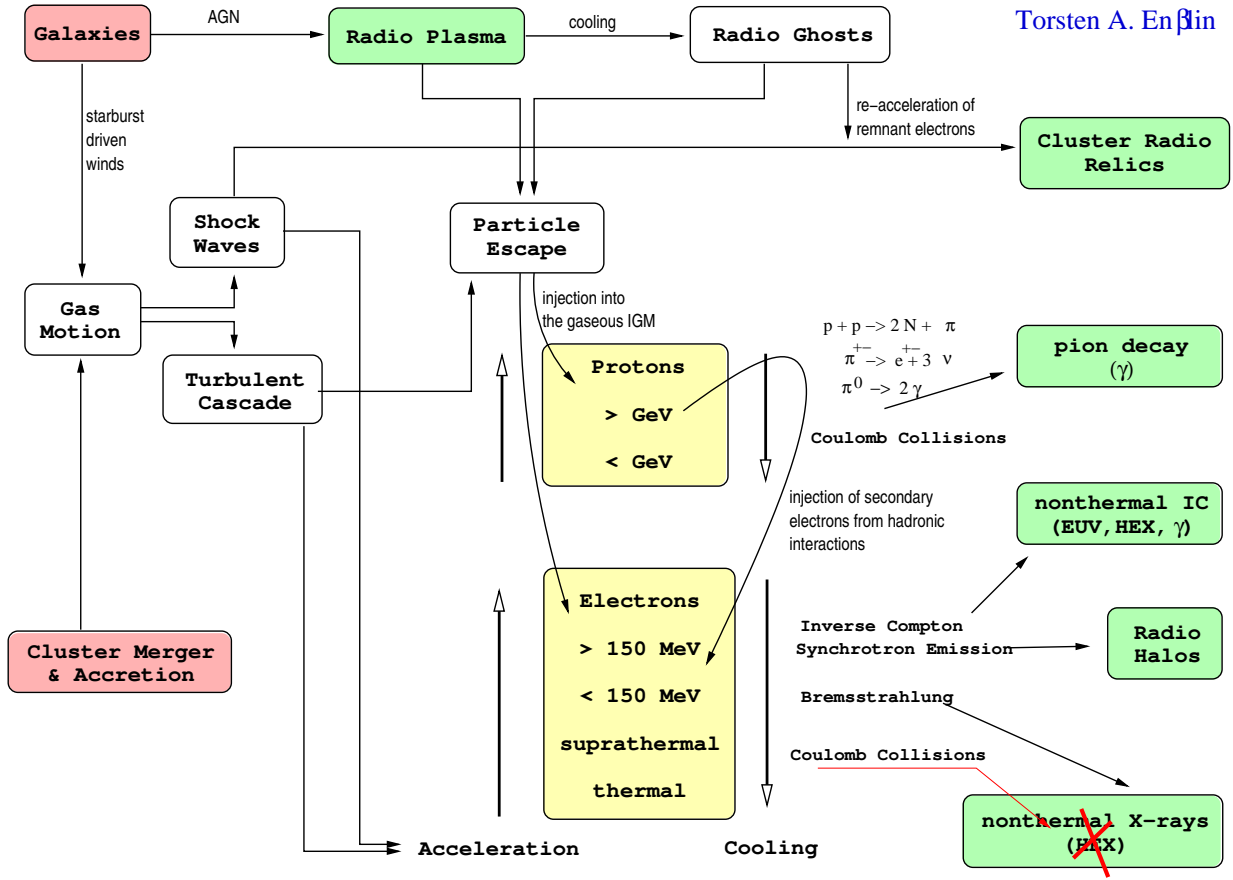}}
%\vskip -1.3cm
\caption{\label{fig:sketch}
Basic theory building blocks covering most of the proposed
scenarios for the production and maintenance of relativistic particle
populations in galaxy clusters and their observational
consequences. Nearly all theoretical scenarios for cluster radio halos
are summarized within this figure (\eg re-acceleration model, hadronic
model, hybrid hadronic-re-acceleration model, and many more). Red boxes:
the main energy sources, yellow boxes: particle populations (CRp,
CRe), green boxes: observables, white boxes: physical
processes/phenomena which are (nearly) unobservable. The red cross
indicates that the possibility that the reported non-thermal X-rays
are generated by Bremsstrahlung of a super-thermal electron population
runs into energetic problems due to strong Coulomb losses of such
electrons (Petrosian 2001).}\NoCite{2001ApJ...557..560P}
%\vskip -0.5cm
\end{figure*}

The existence of ICM magnetic fields and cosmic rays is not too
surprising, since there are plenty of energy sources available,
which could have contributed:
\begin{itemize}
  \item cluster mergers: shock waves and turbulence (\eg Miniati et
  al. 2000, 2001)\NoCite{2000ApJ...542..608M, 2001ApJ...562..233M},
  \item active galactic nuclei (\eg En{\ss}lin \etal 1997,
  1998)\NoCite{1997ApJ...477..560E, 1998AA...333L..47E},
  \item supernovae remnants (\eg V{\"o}lk et
  al. 1996)\NoCite{1996SSRv...75..279V},
  \item galactic wakes (\eg Jaffe 1980, Roland 1981, Ruzmaikin et
  al. 1989)\NoCite{1981A&A....93..407R, 1989MNRAS.241....1R,
  1980ApJ...241..925J},
  \item decaying/annihilating dark matter particles
  (\eg Colafrancesco \& Mele 2001, Boehm et
  al. 2004)\NoCite{2001ApJ...562...24C, 2004JPhG...30..279B}.
\end{itemize}
Furthermore, we have a basic understanding of relevant physical
processes, such as
\begin{itemize}
  \item particle cooling/radiation mechanisms,
  \item particle acceleration by shocks and turbulence, 
  \item magneto-hydrodynamics.
\end{itemize}
In principle, we just have to put all these pieces together in order
to get an understanding of magnetic fields and cosmic rays in galaxy
clusters. This is done for the cosmic rays in a very schematic way in
Fig. \ref{fig:sketch}.  The problem with such a generic approach is,
that there are large uncertainties in the theoretical description of
the different plasma processes which do not allow to decide {\it a
priori} which are the dominant processes are within the complex
network. Here, we will discuss two possibilities:
\begin{itemize}
  \item the relativistic electrons seen in cluster radio halos are
  re-accelerated by cluster turbulence (\eg  Jaffe 1980, Roland 1981,
  Giovannini \etal 1993, Brunetti \etal 2001, 2004, Gitti et
  al. 2001)\NoCite{1993ApJ...406..399G, 2001MNRAS.320..365B,
  2004MNRAS.350.1174B, 2002A&A...386..456G}
  \item the radio emitting CRe are secondaries from hadronic
  interactions of a long-living CRp population within the ICM
  gas. (\eg Dennison 1980, Vestrand 1982, Blasi \& Colafrancesco
  1999, Dolag \& En{\ss}lin 2000, Pfrommer \& En{\ss}lin
  2004a)\NoCite{1980ApJ...239L..93D, 1982AJ.....87.1266V,
  1999APh....12..169B, 2000A&A...362..151D, 2004A&A...413...17P} 
\end{itemize}

\begin {figure*}[t]
\vskip -1.3cm
\centerline{\epsfysize=11.8cm\epsfbox{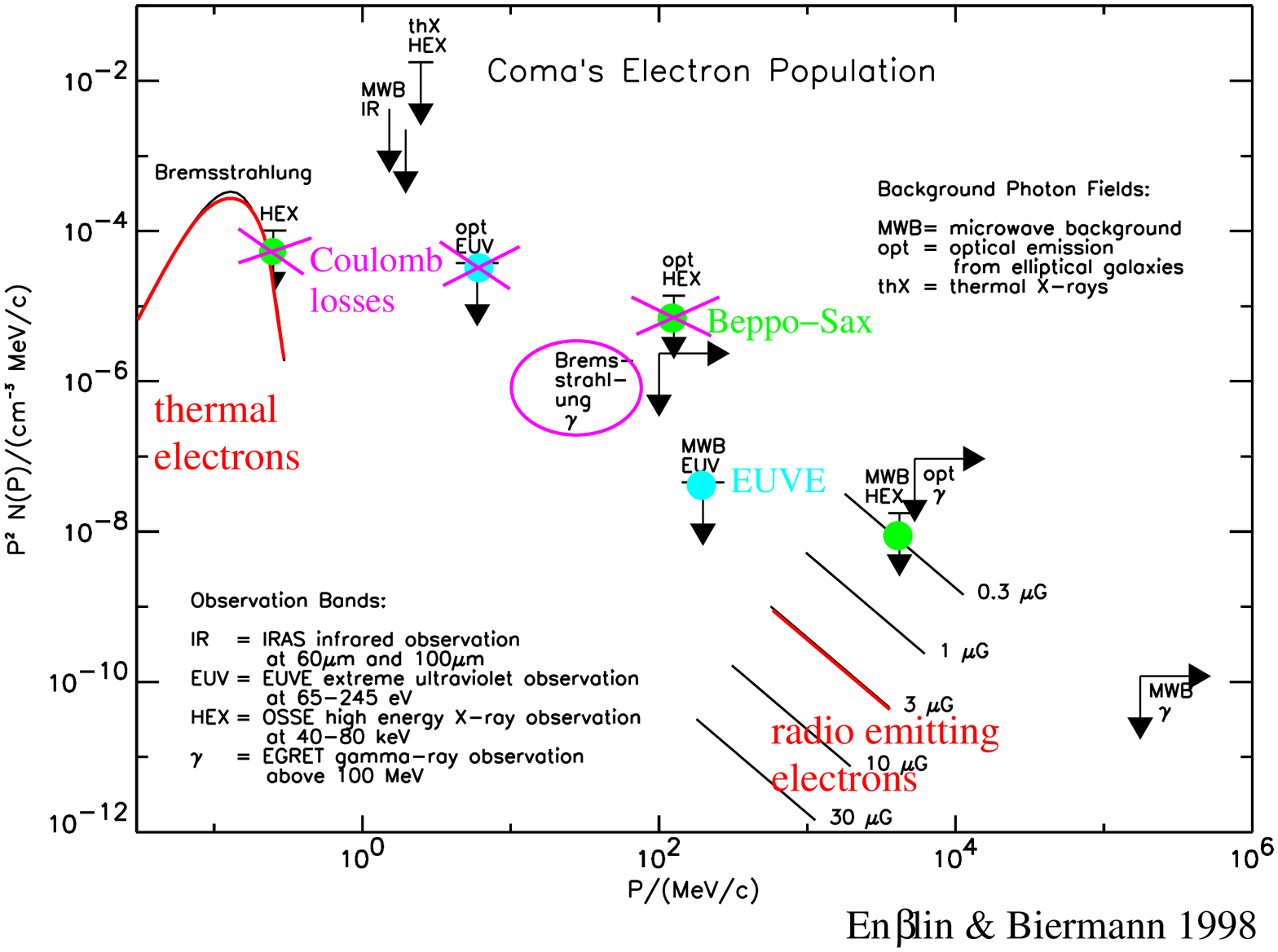}}
\vskip -1.3cm
\caption{\label{fig:espec4}
  Electron spectrum in the Coma cluster of galaxies (adopted
  from En{\ss}lin \& Biermann 1998). Red lines indicate parts of the
  electron spectrum we know to exist: thermal electrons (left), and
  radio-synchrotron electrons (right). The exact spectral
  location of the latter depends on the unknown ICM magnetic field
  strength. The upper limits result from physical processes, which
  would have produced photons in excess of observational constraints,
  if more than the indicated electrons of a given energy were present
  in the ICM. One process is Inverse Compton scattering of known
  background photon fields into observational bands. The other is
  Bremsstrahlung from energetic electrons with the ICM gas. Some of
  the used observations are actual claims of detections, and therfore
  could be measurements of the electron spectrum. Green points
  indicate possible explanations of the reported high energy X-ray
  (HEX) excess measured by Beppo-Sax (Fusco-Femiano \etal 1996,
  2004, but see Rosetti \& Molendi 2004, who do not find a significant
  signal). Blue points indicate possible explanations of the reported
  extreme ultraviolet excess measured by EUVE (\eg Liu et
  al. 1996). However, some of the marked possibilities are unlikely
  (marked by pink crosses): severe Coulomb losses would remove
  super-thermal and mildly relativistic electron populations too
  quickly to be replenished (Petrosian 2001, En{\ss}lin, Lieu,
  Biermann 1999). The IC scattering of optical photons into the HEX
  band would require too many relativistic electrons to be consistent
  with other upper limits resulting from
  Bremsstrahlung.}\NoCite{2001ApJ...557..560P}\NoCite{1998AA...330...90E,
  1999AA...344..409E, 1996Sci...274.1335L, 1999ApJ...513L..21F,
  2004ApJ...602L..73F, 2004A&A...414L..41R}
%\vskip -0.5cm
\end{figure*}

In order to have a chance to discriminate between the different
theoretical models, detailed information on the electron spectrum
would be very helpful, since the different scenarios exhibit different
characteristics. An attempt to collect such information is displayed
in Fig. \ref{fig:espec4}. From the figure and the discussion in the
caption it becomes clear that the interpretation of the Extreme
Ultraviolet (EUV) and High Energy X-ray (HEX) flux reported for Coma
as Inverse Compton scattered CMB photons would require magnetic fields
of $1.4\, \mu$G (EUV) or significantly less (HEX). Such weak field
strengths are an obvious contradiction to the Faraday method based
estimates of several $\mu$G, especially if one compares magnetic
energy densities or electron spectra as plotted in
Fig. \ref{fig:espec4}.

\section {What are our prejudices?}

\subsection{Typical cluster magnetic field strength}

Depending whom one asks, or which paper on cluster physics one consults, the
assumption of the magnetic field strength differ significantly. Sub-micro-Gauss
fields are assumed when an emphasize is given to the Inverse Compton
scattering of CMB photons into the EUV and HEX energy bands. Super-micro-Gauss
fields are assumed in the context of the interpretation of Faraday rotation
signals.

Both methods of field estimates have their weak points. The inverse Compton
argumentation can only provide strict lower limits to magnetic fields
strength. The used observed EUV or HEX flux could have (partially) resulted
from a different source or could be a measurement artefact. Therefore, the
number of relativistic electrons could be smaller than assumed in the
estimate, requiring stronger magnetic fields in order to provide the same
amount of observed synchrotron emission.

Faraday rotation based field estimates are also not straightforward, since
magnetic field reversals along the line of sight partially cancel each other's
Faraday signal. What is left as a typical cluster Faraday signal is the result
of a random walk in rotation measure (RM) in the case of turbulent magnetic
fields. The statistical RM signal depends on the statistical magnetic field
strength times the square root of the magnetic autocorrelation length. The
latter is unknown, and thus the Faraday based field estimates suffer from this
uncertainty.  However, the statistical properties of Faraday maps may allow to
measure the magnetic autocorrelation length under relative reasonable
assumptions of statistical isotropy and homogeneity of the magnetic fields
(see Vogt \& En{\ss}lin, these proceedings).

It can be seen, that the Faraday rotation based field estimates can
not be accommodated by sub-micro-Gauss field strength. For a typical
cluster like Coma, $3-5\, \mu$G are reproducing the Faraday signal if
a magnetic length-scale of 10 kpc is assumed.  Lowering the magnetic
field strength by one order of magnitude -- as suggested by the IC
based field estimates (in the case of the HEX excess) -- would require
an increase of the magnetic length scale to $\sim$ Mpc in order to
reproduce the Faraday signal strength. But fields ordered on the
cluster size would produce a nearly homogeneous RM signal, and not
exhibit the many sign reversals observed in clusters.

\subsection{The origin of radio halos}

Also the origin of the relativistic electrons producing the cluster radio
halos is controversially debated. There are the re-acceleration models, which
are favored by some authors since they are able to reproduce nearly every
observational feature reported so far. And then there are the hadronic models,
which are favored by others not only because of their simplicity, but also
since the necessary cosmic ray proton population should be present in the ICM
because most cosmic ray acceleration mechanisms are believed to preferentially
accelerate protons.

\section {How to verify models?}

It is often claimed that a model or theory is verified. However, theories can
only be falsified.  To be a scientific theory, it has to be falsifiable
(Popper). This means, {\it it must be possible to derive from it unambiguous
predictions for doable experiments such that, were contrary results be found,
at least one premise of the theory would have proven not to apply to nature.}

Let us compare the different models for radio halos in the light of this
statement:

\subsection{Hadronic model}
The hadronic model is therefore a very good theory in the sense of
Popper, since it makes a number of testable predictions. 
\begin{enumerate}
\item The hadronic model predicts gamma and neutrino fluxes from clusters of
galaxies.The gamma ray flux should be detectable with GLAST (Pfrommer \&
En{\ss}lin and also Reimer, these proceedings),
\item The hadronic model can not explain very strong spectral bending in the
  radio, since even a mono-energetic proton population injects a broad
  electron spectrum, which produces an even broader synchrotron spectrum.
\item Furthermore, the necessary energy budget of the hadronic model can
 exceed the available energy sources, especially in the outer, low-density
 regions of clusters, where targets for an efficient CRp to CRe conversion are
 rare.
\end{enumerate}
Some of these predictions seem to put already some stress on the hadronic
model.

There is a strong spectral break reported for the total flux of the Coma
cluster (Thierbach \etal 2003\NoCite{2003A&A...397...53T}), which would be too strong for the
hadronic model. Actually, this spectral break is too strong for any kind of
synchrotron emission, indicating that there might be some problems with the
spectral dataset of the Coma radio halo compiled from the literature. The
strong spectral break might be partly explained by a contamination of the high
frequency measurements by the Sunyaev-Zeldovich effect, however, it could also
be an artefact of the very difficult radio astronomical measurement of
extended radio halo luminosities in the vicinity of very strong point
sources.

It was shown that at least the radio halos in the Coma and Perseus
clusters do not violate the energy constraints (Pfrommer \& En{\ss}lin
2004b)\NoCite{2004MNRAS.352...76P}. In order to produce the observed
radio emission without too much energy in CRp, the hadronic models
favor larger magnetic field strength, in the range which are
supported by the Faraday rotation measurements.

The prediction of gamma rays could have not been tested by existing telescope
sensitivities. Future gamma ray telescopes have the potential to ultimately
refute the hadronic model, since there is an unavoidable minimum gamma ray
flux of the order of the total radio flux predicted for this class of
models. If magnetic field strength are as low as $3\, \mu$G or less, the
predicted gamma ray flux of the hadronic model has to be larger than this
minimum prediction.

\subsection{Re-acceleration models}

The re-acceleration model has not made unique predictions which have the
potential to allow to refute the model observationally. The model seems to be
able to fit any yet reported radio profile and spectra due to its many poorly
constrained parameters. Therefore, distinctive predictions of the
re-acceleration model would be very important, if they can be made at all (see
Brunetti in these proceedings for first predictions of the re-acceleration
model).

\section {How to access uncertainties?}

Reliable spectral information on radio halos has the potential to refute the
hadronic model. However, there are many known, and maybe some unknown sources
of artefacts in radio astronomy, which can perturb the measured spectrum.  To
understand the meaning and significance of features and spectra we need an
end-to-end analysis of the data reduction process. This analysis should go
from the detector signal, through (self)-calibration, to the map making, point
source removal and halo flux integration step. Without such an analysis it
would be premature to claim that the hadronic model is ruled out due to the
reported spectral steepening of the Coma radio halo.

Also for the Faraday rotation measurements such an analysis would be highly
desirable, in order to understand if and how observational artefacts influence
our field estimates. In order to go into this direction, methods to quantify
the level of noise and artefacts in Faraday maps were developed (En{\ss}lin
\etal 2003). They were used to verify the improved quality of Faraday maps and
even the accuracy of Faraday error maps generated with the new {\it
Polarization Angle Correcting rotation Measure Analysis} (PACMAN) algorithm
(Dolag et al. 2004, Vogt \etal 2004)\NoCite{2004astro.ph..1214D,
2004astro.ph..1216V}. These maps were then analyzed with a maximum likelihood
power spectrum estimator (Vogt \& En{\ss}lin, submitted), which is based on
the cross correlation of Faraday signals in pixel pairs, as expected for a
given magnetic power spectrum and galaxy cluster geometry (En{\ss}lin \& Vogt
2003, Vogt \& En{\ss}lin 2003)\NoCite{2003A&A...401..835E,
2003A&A...412..373V}.

\begin {figure}[t]
%\vskip -1.3cm
\centerline{\epsfysize=0.3\textwidth\epsfbox{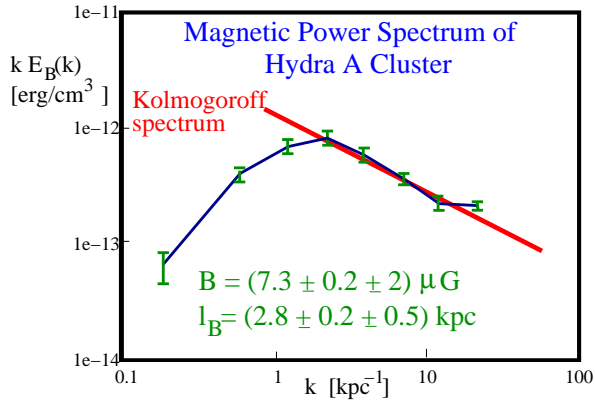}}
%\vskip -1.3cm
\caption{\label{fig:magpow}
Magnetic power spectrum within the cooling flow region of the
  Hydra A cluster of galaxies (for details see Vogt \& En{\ss}lin,
  these proceedings). The last data-point at large k-values is likely
  affected by map noise and should be discarded.}
%\vskip -0.5cm
\end{figure}

The result of this exercise is not only a magnetic power spectrum, which is
corrected for the complicated geometry of the used radio galaxy and of the
Faraday screen, but also an assessment of the errors, and even the cross
correlation of the errors. The power spectrum of the Hydra A cluster cool core
region, which is displayed in Fig. \ref{fig:magpow}, exhibits a
Kolmogoroff-like power law on small scales, a concentration of magnetic power
on a scale of 3 kpc (the magnetic auto-correlation length) and a total field
strength of $7\pm2 \,\mu$G. The given error is the systematic error due to
uncertainties in the Faraday screen geometry. The statistical error is lower
by one order of magnitude.

\section {Is a consistent picture possible?}
\subsection {Combining observational information}

Here, I am attempting to draw a consistent picture, which may explain
at least a significant subset of the observational
information\footnote{There are obviously some reported observations,
which are not included in this subset: the HEX excess, since it's
significance is debated; the reported spectral cut-off of the Coma
radio halo emission, which may or may not be a combination of
observational artefacts of the difficile radio observation and data
reduction plus some impact of the thermal Sunyaev-Zeldovich
decrement.}:
\begin{itemize}
  \item the Faraday rotation observations, which point towards turbulent
  fields strength of several $\mu$G strength: in the cool core region of the
  Hydra A cluster a field strength of 7$\mu$G correlated on 3 kpc; in
  non-cooling flow clusters like Coma somewhat lower fields (say 3$\mu$G) with
  a somewhat larger correlation length (say 10-30 kpc). 
  \item the radio halo synchrotron emission of CRe in Coma 
  \item the EUV excess of the Coma cluster, which may be understood as being
  inverse Compton scattered CMB light
\end{itemize}
The Faraday measurements provide us with volume averaged\footnote{Actually,
  it is more an average with the thermal electron distribution. However, the
  latter varies slowly on scales over which the magnetic fields vary, so that
  it can be factorized out of the signal.} magnetic energy
densities, since the RM dispersion scales as
\begin{equation}
  \langle {\rm RM}^2 \rangle \propto   \langle B^2 \rangle_{\rm
  Vol}.
\end{equation}
The synchrotron emission is also (approximately) proportional to the magnetic
energy density, but weighted with the CRe population around
10 GeV:
\begin{equation}
  L_{\rm radio} \propto   \langle B^2 \, n_{\rm CRe}\rangle_{\rm Vol}.
\end{equation}
Finally, the inverse Compton flux is a direct measurement of the number
density of CRe (of the appropriate energy, see
Fig. \ref{fig:espec4}):
\begin{equation}
  L_{\rm IC} \propto   \langle n_{\rm CRe}\rangle_{\rm Vol}.
\end{equation}
Combining the latter two measurements provides a magnetic field estimate (for
a given or assumed electron spectral slope), which is weighted with the CRe
density:
\begin{equation}
\langle B^2 \rangle_{\rm CRe} =   \frac{\langle B^2 \, n_{\rm
    CRe}\rangle_{\rm Vol}}{\langle n_{\rm CRe}\rangle_{\rm
    Vol}} \propto  \frac{L_{\rm radio}}{L_{\rm IC}}
\end{equation}
\subsection {Reconciling the discrepant field estimates}

The magnetic energy density derived from the combination of
synchrotron and inverse Compton flux is at least one order of
magnitude lower than the one derived from RM measurements. This
discrepancy might be reconciled if there is a significant difference
between volume and CRe weighted averages. This would require
\begin{enumerate}
  \item an inhomogeneous magnetic energy distribution,
    \item an inhomogeneous distribution of the CRe,
      \item an anti-correlation between these two.
\end{enumerate}
The required conditions could be generated if there are physical mechanisms
which 
\begin{itemize}
  \item[a.] produce inhomogeneous or intermittent magnetic fields
  \item[b.] anti-correlate the CRe density with respect to the
  magnetic energy density.
\end{itemize}
A very natural physical mechanism, which could provide this anti-correlation
is synchrotron cooling in inhomogeneous magnetic fields. In case of an
injection rate of CRe which is un-correlated with the field strength, the
equilibrium electron density is
\begin{equation}
  n_{\rm CRe} \propto (B^2 + B^2_{\rm CMB})^{-1},
\end{equation}
where $B_{\rm CMB} \approx 3.2 \mu$G describes the field strength
equivalent to the CMB energy density. 

For illustration, we assume that only a small fraction $f_B = 0.1$ of
the volume is significantly magnetized with a field strength of
$10\,\mu$G, and the rest with only $1 \,\mu$G. We will see later that
$f_B = 0.1$ may be a plausible number.  The volume average would give
$\langle B^2 \rangle_{\rm Vol}^{1/2} = 3.3\, \mu$G, whereas the CRe
average gives $\langle B^2 \rangle_{\rm CRe}^{1/2} = 1.5\,
\mu$G. These numbers are in good agreement with the corresponding
field estimates for the Coma cluster based on Faraday rotation and
IC/synchrotron measurements, respectively. A larger ratio in magnetic
field estimates could even be accomodated since the EUV emitting
electrons are at energies below the synchrotron electrons. A spectral
bump of an old accumulated electron population at these energies is
therefore possible, and even plausible due to the minimum in the
electron cooling rate at these energies.

The hadronic model would provide a CRe injection rate which is not correlated
with the magnetic field strength, as would be required by the above
explanation of the discrepancy of magnetic field estimates by the two methods
would work. In contrast to this, in the re-acceleration model one would expect
a strong positive correlation of CRe and magnetic field strength, since
magnetic fields are essential for the CRe acceleration.

\subsection {Turbulent magnetic dynamo theory}

It remains to be shown that there is also a natural mechanism
producing intermittent magnetic fields. The Kolmogoroff-like magnetic
power spectrum in the cool core of the Hydra A cluster indicates that
the magnetic fields are shaped and probably amplified by
hydrodynamical turbulence (\eg De Young
1992)\NoCite{1992ApJ...386..464D}. Therefore, we have to look into the
predictions of the theories of turbulent dynamo theories.

It is generally found by a number of researchers that the non-helical
turbulent dynamo saturates in a state with a characteristic magnetic
field spectrum (\eg Ruzmaikin \etal 1989, Sokolov \etal 1990,
Subramanian 1999 and many others)\NoCite{1989MNRAS.241....1R,
1990IAUS..140..499S, 1999PhRvL..83.2957S}. The effective magnetic
Reynolds number (including turbulent diffusion) reaches a critical
value of $R_{\rm c} \approx $ 20 ... 60.  The magnetic fields should
exhibit -- more or less pronounced -- the following properties:
\begin{itemize}
\item[A.] The
average magnetic energy density $\varepsilon_B$ is lower than the turbulent kinetic energy
density $\varepsilon_{\rm kin}$ by $\varepsilon_B \approx \varepsilon_{\rm
  kin} \, R_{\rm c}^{-1}$. 
\item[B.] The magnetic fluctuations are concentrated on a scale $l$, which
is smaller than the hydrodynamical turbulence injection scale $L$ by $l
\approx L R_{\rm c}^{-1/2}$.
\item[C.] Correlations exist up to scale $L$, turn there into an
  anti-correlation, and quickly decay on larger scales.
\item[D.] This may be understood by Zeldovich's flux rope model, in which
  magnetic ropes with diameter $l$ are bent on scales of the order $L$.
\item[E.] Within flux ropes, magnetic fields can be in equipartition with the
  average turbulent kinetic energy density. 
\item[F.] The magnetic drag of such ropes produces a hydrodynamical viscosity
  on large scales, which is of the order of $4\%$ of the turbulent
  diffusivity (Longcope \etal 2003).
\end{itemize}

\subsection {Confronting theory with observations}

Turbulent magnetic dynamo theory predicts intermittent magnetic
fields, as favored by the proposed explanation of the discrepancy in
the different magnetic field estimate methods. Let's see if the other
predictions of the theory are in agreement with observations. We
assume, that $R_{\rm c}$ is in the range 20 to 60.

\begin{itemize}
\item[A.] The expected turbulent energy density in the Hydra A cluster
  core is of the order of $(0.3 \ldots 1)\,
  10^{-10}\, {\rm erg\, cm^{-3}}$, which corresponds to turbulent
  velocities of $v_{\rm turb} \approx (300 \ldots 500) \,{\rm km/s}$. This
  is comparable to velocities of buoyant radio plasma bubbles (En{\ss}lin \&
  Heinz 2002)\NoCite{2002A&A...384L..27E}, which are expected to stir
  up turbulence (\eg Churazov \etal 2001)\NoCite{2001ApJ...554..261C}.
\item[B.] The expected turbulence injection scale in the Hydra A
  cluster core is of the order of $(15 \ldots 25)$ kpc, again consistent with
  the radio plasma of Hydra A being the source of turbulence. The
  dynamical connection of the radio source length scale and the magnetic
  turbulence scale would explain why the Faraday map of Hydra A is
  conveniently sized to show us the peak of the magnetic power spectrum.
\item[C.] This prediction is hard to test since fluctuations on scales
  larger than the radio source are not measurable due to the limited
  RM map size. However, the downturn of the magnetic power spectrum at
  small $k$-values is at least in agreement with this prediction.
\item[D.] Magnetic intermittency in form of flux ropes might have been
  detected as stripy patterns in the RM map of 3C465 (Eilek \& Owen
  2002)\NoCite{2002ApJ...567..202E}.
\item[E.] The fraction of strongly magnetized volume can become as
  small as $f_{\rm B}  = R_{\rm c}^{-1} \approx 0.02 ... 0.05$, a value which is
  more extreme than what we assumed in our example for the Coma cluster.
\item[F.] The expected hydrodynamical viscosity on large scales in the Hydra
  cluster is of the order of $(1\ldots 4) \, 10^{28}\,{\rm cm^2/s}$. It is
  interesting to note, that a lower limit on the large scale viscosity of the
  comparable Perseus cluster cool core of $4\cdot 10^{27}\, {\rm cm^2/s}$ was
  estimated by Fabian et al. (2003)\NoCite{2003MNRAS.344L..48F}. An upper
  limit on the viscosity in the (somewhat different) Coma cluster of $\sim
  3\cdot \,10^{29}\, {\rm cm^2/s}$ was derived by Sch{\"u}cker \etal
  (2004)\NoCite{2004A&A...426..387S}. Both limits are consistent with our
  coarse estimate of the large scale viscosity and enclose it.
\end{itemize}

\section{Conclusion}

To summarize, it is not clear if all observational information can be
fitted into a single consistent picture of extragalactic cosmic rays
and magnetic fields. It might be that we have to revise some of our
data-points. The choice of data-points which were used (or ignored)
here for the construction of the presented physical picture reflects
the author's personal prejudices and may certainly be questioned.

Nevertheless, it should have become clear that the existence of strong and
possibly intermittent magnetic fields (several $\mu$G) in galaxy clusters is
strongly supported by the recent detection of a Kolmogoroff-like magnetic
power spectra in the Hydra A cluster. Some of the discrepancies between
Faraday-based and inverse Compton-based field estimates can be explained by
effects caused by magnetic intermittence, which is expected from turbulent
dynamo theory.

Furthermore, it is argued that the hadronic generation mechanism of the
cluster radio halo emitting electrons is a viable model. This model is
providing a number of stringent predictions (like minimal gamma ray fluxes,
limits on spectral bending, maximal possible radio luminosities), which allows
detailed consistency tests with future sensitive measurements. The fact that
this class of models seems to be more under pressure by some reported
observations compared to alternatives as the also viable re-acceleration model
may be caused by the larger number of predictions of the hadronic model in
conjunction with underestimated systematic observational errors.

\acknowledgements{It is a pleasure to thank the conference organizers
  for the excellent meeting, and the very warm hospitality. I also
  want to thank Corina Vogt and Christoph Pfrommer for discussion and
  comments on the manuscript}

%\bibliography{aamnem99,tae}
%\bibliographystyle{aabib99}

\end{document}